\documentstyle[12pt,dina4,psfig,amssymb,psfig]{article}
\begin{document}
\parindent=0cm
\parskip=1.5mm

\def\bi{\begin{list}{$\bullet$}{\parsep=0.5\baselineskip
\topsep=\parsep \itemsep=0pt}}
\def\ei{\end{list}}
\def\lusa#1{({\bf{#1}})\marginpar{$\longleftarrow$ LuSa?}}
\def\as#1{({\bf{#1}})\marginpar{$\longleftarrow$ AS?}}

\def\phi{\varphi}
\def\-{{\bf --}}
\def\vm{v_{max}}
\def\ta{\tilde a}
\def\ts{\tilde s}
\def\tr{\tilde r}
\newcommand{\s}{\sigma}
\newcommand{\la}{\lambda}
\newcommand{\eps}{\varepsilon}
\newcommand{\al}{\alpha}
\newcommand{\nn}{{\cal N}}

\newcommand{\be}{\begin{equation}}
\newcommand{\ee}{\end{equation}}
\newcommand{\bea}{\begin{eqnarray}}
\newcommand{\eea}{\end{eqnarray}}

\newcommand{\nonu}{\nonumber\\}

\renewcommand{\thefootnote}{\fnsymbol{footnote}}

\parindent=0cm
\begin{center}
  {\LARGE\bf Metastable States in Cellular Automata}\\[0.8cm]
\end{center}
\begin{center}
  {\LARGE\bf  for Traffic Flow}
\end{center}
\vskip1.8cm 
\setcounter{footnote}{1}
\begin{center}
  {\Large R.\ Barlovic$^{1}$, L.\ Santen$^{2}$, A.\ Schadschneider$^{2}$
and M.\ Schreckenberg$^{1}$}
\end{center}
\vskip1.3cm
\begin{center}
  $^{1}$ Theoretische Physik/FB 10\\ 
  Gerhard-Mercator-Universit\"at Duisburg\\ 
  D--47048 Duisburg, Germany\\
  email: {\tt barlovic,schreck@uni-duisburg.de}
\end{center}
\begin{center}
  $^{2}$  Institut f\"ur Theoretische Physik\\ Universit\"at zu K\"oln\\ 
  D--50937 K\"oln, Germany\\
  email: {\tt santen,as@thp.uni-koeln.de} 
\end{center}
\begin{center}
\today\\[1cm]
\end{center}

\begin{center}
{\em Dedicated to J.\ Zittartz on the occasion of his 60th birthday}\\[1cm]
\end{center}

\vskip1.5cm {\large \bf Abstract}\\[0.2cm]
\hspace{.4cm} 
Measurements on real traffic have revealed the existence of metastable 
states with very high flow. Such states have not been observed in the
Nagel-Schreckenberg (NaSch) model which is the basic cellular automaton for
the description of traffic. Here we propose a simple generalization of
the NaSch model by introducing a velocity-dependent randomization. We
investigate a special case which belongs to the so-called slow-to-start
rules. It is shown that this model exhibits metastable states, thus sheding
some light on the prerequisites for the 
occurance of hysteresis effects in the flow-density relation.
\vskip2cm
\vfill
\pagebreak
\renewcommand{\thefootnote}{\arabic{footnote}}
\setcounter{footnote}{0}
\section{Introduction}

A few years ago, Nagel and Schreckenberg \cite{NaSch} (NaSch) have proposed a 
probabilistic cellular automaton (CA) for the description of 
single-lane highway traffic\footnote{For an overview of other approaches, see 
e.g.\ \cite{juelich,duis,helbing,may}}.
Using very simple rules, this model is able to reproduce the basic
phenomena encountered in real traffic, e.g.\ the occurance of phantom
traffic jams ('jams out of nowhere'). The NaSch model is 'minimal' in the
sense that every simplification of the rules no longer produces realistic
results. On the other hand, for the description of more complex situations
(e.g.\ multi-lane traffic, ramps)\footnote{For applications to urban traffic, 
see e.g.\ \cite{dallas,duisburg}.} or for a proper modelling of the
'fine-structure' of traffic flow, additional rules have to be added and/or
the basic rules have to be modified.

The NaSch model \cite{NaSch} is a probabilistic cellular automaton. 
Space and time (and hence the velocities) are discrete. The road is divided 
into cells of length 7.5 m. Each cell can either be empty or occupied by 
just one car. The state
of car $j$ ($j=1,\ldots,N$) is characterised by its momentary velocity 
$v_j$ ($v_j=0,1,\ldots, \vm$). The state of the system at time $t+1$ can be
obtained from the state at time $t$ by applying the following four 
rules to all cars at the same time (parallel dynamics):
\begin{description}
\item[R1:] Acceleration:\ \ \  $v_j(t)\rightarrow v_j(t+\frac{1}{3})
=\min\{v_j(t)+1,\vm\}$
\item[R2:] Braking:\ \ \  if\ \  $v_j(t+\frac{1}{3}) > d_j(t)$\ \ \ then\ \ \  
$v_j(t+\frac{2}{3})=d_j(t)$\\
\phantom{ Braking:\ \ \  if\ \  $v_j(t+\frac{1}{3}) > d_j(t)$\ \,}else\ \ \ \ 
$v_j(t+\frac{2}{3})=v_j(t+\frac{1}{3})$
\item[R3:] Randomization:\ \ \  $v_j(t+\frac{2}{3})\ 
{\stackrel{p}{\rightarrow}}\ v_j(t+1)=\max\{0,v_j(t+\frac{2}{3})-1\}$\\
\phantom{ Randomization:} with probability $p$ 
\item[R4:] Driving:\ \ \  car $j$ moves $v_j(t+1)$ cells.
\end{description}
Here $d_j(t)$ denotes the number of empty cells in front of car $j$, i.e.\ 
the gap or headway. One timestep $t\to t+1$
corresponds to approximately 1 sec in real time \cite{NaSch}. 

In the spirit of modelling complex phenomena in statistical physics,
the NaSch model does not try to describe traffic flow very accurately
on a microscopic level. Macroscopic effects observed in real traffic,
e.g.\ the spontaneous formation of jams, can be understood by
introducing just one simple stochastic parameter, the braking
probability $p$. Note that the motion of a single car might exhibit
(on short timescales) unrealistic features, like stopping from maximum
velocity within a few timesteps without any reason. However, after
averaging over the motion of all cars or on long timescales, the NaSch
model produces quite realistic results. Therefore one should not try
to relate these large fluctuations to those observed in real traffic.

Besides the CA models which are discrete in space and time, several
other approaches to traffic flow have been discussed recently. Among
these are space-continuous models in discrete time like the model of
Krauss et.\ al.\ \cite{zprmod1,zprmod2}, as well as models continuous both 
in space and time, e.g.\ the macroscopic (fluid-dynamical) models \cite{KK}, 
the optimum-velocity model \cite{bando}, coupled-map models \cite{kikuchi} 
and gas-kinetic models \cite{helbinggas}. For further
references we refer to \cite{juelich,duis,helbing}.

In the present paper we want to investigate hysteresis effects encountered
in empirical observations \cite{Treit,Kerner2,helbingexp}. Such effects are
related to the existence of metastable states in certain density regimes. In
the NaSch model these states have not be observed. Here we want to present
slightly modified models which are able to produce metastable states and
hysteresis. As a consequence, the occurance of metastable states is
therefore not related to the use of realistic braking rules, a continuum
description or deterministic models.

In order to establish the existence of metastable states one can follow
two basic strategies. In the first method, the density of cars is changed 
adiabatically by adding or removing vehicles from the stationary
state at a certain density. Starting in the jamming phase (large densities)
and removing cars, one obtains the lower branch of the hysteresis curve.
On the other hand, by adding cars to a free flowing state (low densities),
one obtains the upper branch. The second method does not require changing
the density. Instead one starts from two different initial conditions, the
megajam and the homogeneous state. The megajam consists of one large, compact
cluster of standing cars. In the homogeneous state, cars are distributed
equidistantly (with one large gap for incommensurate densities). 

In certain density regimes the fundamental diagram can consist of two
branches. In the upper branch (with higher flow) there are almost
no interactions between the cars and the system remains in a homogeneous,
jam-free state.
In the lower branch, however, the system is in a `phase-separated' state,
consisting of one large jam and a free-flowing part. 

Experimental observations \cite{KernerReh2} suggest that a reduction
of the outflow from a jam compared to the maximum possible
flow\footnote{This reduction is often referred to as `capacity drop'.}
is an important ingredient for the occurrance of metastable states with
large lifetimes. Such a reduction can be implemented by so-called
slow-to-start rules, where standing cars accelerate with lower
probability than moving cars.  It leads to a downstream flow with a
relatively low density corresponding to the lower branch of the
hysteresis curve.  Due to the reduction of the density in the outflow
region of a jam, the flow in the jammed state is significantly lower
than the flow of the homogeneous state at the same density.

The paper is organized as follows: In Sect.\ \ref{sec_s2s} two models
with slow-to-start rules are investigated. In Sect.\ \ref{sec_nasch} 
we introduce a slightly modified NaSch model with a velocity-dependent 
randomization. In this way, one incorporates the
basic ingredients necessary to produce hysteresis.
In the concluding Section \ref{sec_disc} we summarize our results and
compare with those for other models and real traffic.

\section{Models with slow-to-start rules}
\label{sec_s2s}

In this Section we briefly present results for two models with slow-to-start
(s2s) rules \cite{slow_comf}, the T$^{2}$ model \cite{T2mod,fukui} and the 
BJH model \cite{BJH}.
These models have been introduced in order to model the restart behaviour 
of stopped cars in a more realistic fashion. As will be demonstrated below,
these s2s rules are an important ingredient for the occurance of metastable
states, although the authors of \cite{T2mod,BJH} did not realize the
connection between s2s rules and metastability encountered in real traffic.

\subsection{T$^{2}$ model}

Takayasu and Takayasu (T$^{2}$) \cite{T2mod} have first suggested a CA
model with a s2s rule. This rule has been generalized in \cite{slow_comf}
as follows: A standing car with exactly one empty cell in front of it 
accelerates with probability $q_t = 1-p_t$, while all other cars
accelerate deterministically. The other update rules (R2-R4) of the 
NaSch model are unchanged, e.g.\ all cars are still subject to the
randomization step.
Due to this modification already for $v_{max}=1$ the
particle-hole symmetry is  broken.

\begin{figure}[h]
 \centerline{\psfig{figure=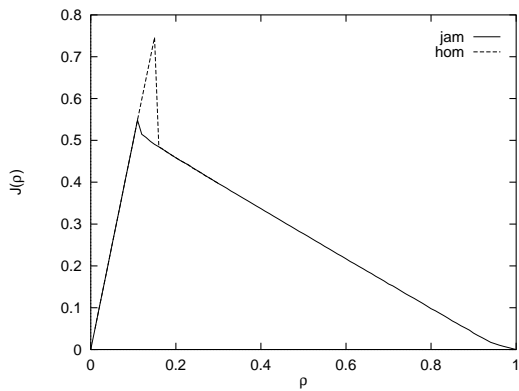,bbllx=50pt,bblly=50pt,bburx=550pt,bbury=400pt,height=5cm}}
\caption{\protect{Fundamental diagram of the T$^2$ model ($v_{max}= 5$,
   $p = 0.01$, $p_t =0.75$) obtained by starting from two different 
   initial conditions, a completely jammed state (full line) and a 
   homogeneous state (broken line). The peak of the full line
   is a finite-size effect.}}
\label{taka_fund}
\end{figure}

In Fig.~\ref{taka_fund} we show the fundamental diagram of the T$^2$
model with $v_{max} =5$, $p=0.01$ and $p_t =0.75$. The system size
used for the simulation was $L=1000$. In order to equilibrate the
system, 10000 lattice updates have been performed. The data shown in 
Fig.~\ref{taka_fund} represent an average over 100000 sweeps through 
the lattice.

Comparing the simulation results with those for the NaSch model,
two qualitative differences have to be discussed.
First, the fundamental diagram has an inflection point
in the high density regime and second the flow behaviour is non-unique 
in a density regime below the density of maximum flow $\rho_{max}$. 

The existence of an inflection point for large $p_t$ has first been 
noticed in \cite{slow_comf} for the case $\vm=1$. In contrast, the 
fundamental diagram of the NaSch model is always convex. The 
different behaviour of the T$^2$ model for large $p_t$ and large
densities $\rho$ is related to the fact that space is not used as 
efficiently as in the NaSch model. One finds a similar behaviour in 
space-continuous models \cite{Krauss}.
There is some experimental evidence that in certain situations the shape of
the fundamental diagram differs from the convex form. This
behaviour of the average flow can be easily obtained tuning the parameter
$p_t$ \cite{slow_comf,duiproc}.

The non-unique behaviour of the flow for densities just below $\rho_{max}$
is due to the fact that for these densities the average flow still 
depends on the initial configuration. The measurements in Fig.~\ref{taka_fund}
have been performed by applying the second method described in the
introduction. The lower branch of the fundamental diagram corresponds
to measurements starting from a initially completely jammed configuration 
while the upper branch  has been obtained starting with a homogeneous 
initialization. It should be mentioned that the same result can also be 
obtained by applying the first method, i.e.\ by changing the density
adiabatically. Since the flow depends on the history of the system,
diagrams like those of Fig.~\ref{taka_fund} are usually called hysteresis
curves.

For low densities the stationary state consists of homogeneous configurations
which are completely jam-free. At higher values of the global density 
the configurations contain one large jam. 
In contrast to the NaSch model, no spontaneous formation of jams in the 
outflow region of the large jam has been observed for the system sizes 
we took into account. Therefore the jammed states are phase separated 
states, unlike for the case of the NaSch model.

It should be noted that the maximum value of the average flow still
depends on the number of updates as well as on the system
size. Therefore we can not exclude that the stationary value of the
average flow is unique in the thermodynamic limit. Nevertherless
the metastable states are extremely stable, even for large systems.
Moreover, in contrast to standard problems of statistical mechanics,
not the thermodynamic limit is relevant for practical purposes, but 
the behavior of systems of finite length (note that 10000 lattice sites 
correspond to a road of length 75~km in reality).
Since in reality roads and observation times are always finite, 
the above results are sufficient for all practical purposes.

\begin{figure}[h]
 \centerline{\psfig{figure=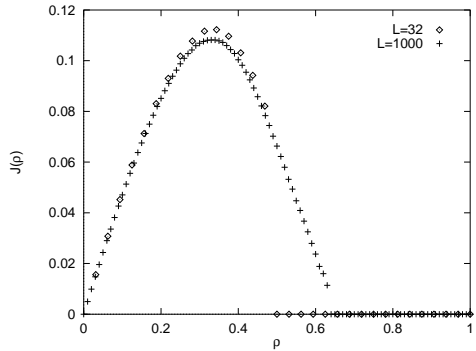,bbllx=40pt,bblly=35pt,bburx=540pt,bbury=420pt,height=5cm}}
\caption{\protect{Fundamental diagram of the T$^2$ model ($v_{max}= 1$,
   $p = 0.5$, $p_t =1$) for two system sizes. For densities slightly
  above $\rho = 0.5$ the stationary state could only be reached for
  the smaller system.}}
\label{T2_fund}
\end{figure}

In Fig.~\ref{T2_fund} we show the fundamental diagram for
$v_{max}=1$, $p=0.5$ and $p_t =1$, i.e. stopped cars can only
move if there are at least two empty cells in front.  Obviously
completely blocked states exist for densities $\rho \geq 0.5$, where
the number of empty cells in front of each car is smaller than
two. Since fluctuations are absent in those states, they have an 
infinite lifetime. Therefore the flow in the stationary state is zero.
In the region $0.5\le \rho \lesssim 0.66$ states with a finite flow exist.
Although these states are not stationary, one has to
perform an extremely large number of update steps until the flow
vanishes for large system sizes and densities  slightly above $\rho = 0.5$, 
because the number of blocked configurations is very 
small compared to the total number of configurations. Precisely at
$\rho = 0.5$, the blocked state is unique and the typical time to
reach this state diverges exponentially with the system size. Therefore
we used very small systems in order to obtain the lower branch of
the fundamental diagram. 

Note that the mechanism for metastability in the case $p_t=1$ is different.
The hysteresis curve in Fig.~\ref{T2_fund} has been obtained by starting
from two different {\em homogeneous} states, differing only in the
velocity of the cars. The jammed branch is reached by starting with a
configuration where all cars $j$ have a velocity $v_j(t=0)=0$, whereas
the upper branch corresponds to an initialization $v_j(t=0)=\vm$. 
In contrast, by following the first method described in the introduction,
one would not find hysteresis, since a megajam initialization would
also yield the high-current branch.

The reason for the occurance of hysteresis in the limit $\vm=1$ and
$p_t=1$ is the existence of a ``geometrical'' phase transition. If
the critical density is exceeded, the cars can no longer move since there is
not enough free space. The ``geometrical'' phase transition
makes it possible to find hysteresis even in the case $\vm=1$.

Finally we want to remark that the $p_t =1$ limit of the T$^2$ model is
some sense complementary to the cruise-control limit \cite{Pac} of the 
NaSch model. In the T$^2$ model the completely blocked state is stabilized
due to the absence of fluctuations whereas in the cruise-control limit
one finds the absence of fluctuations for homogeneous states at low
densities.

\subsection{BJH model}

The s2s rule of the T$^2$ model is a `spatial' rule. The range of
interaction for standing cars is larger than in the NaSch model and
the restart behaviour depends only on the spatial arrangement of the vehicles.
However, there are other ways of implementing a s2s behaviour.
In the Benjamin-Johnson-Hui (BJH) model \cite{BJH} cars which had to
brake due to the next car ahead, will move on the next opportunity only
with probability $1-p_s$. Note that in contrast to the T$^2$ this slow-to-start
rule requires `memory', i.e.\ it is a `temporal' rule depending on the 
number of trials and not on the free space available in front of the car.

For the BJH model no metastable states and hysteresis effects have been 
found until now since only the case $\vm=1$ has been investigated
thoroughly \cite{BJH,slow_comf}. For higher velocities we expect the 
occurance of metastable states, since also in the BJH model the outflow 
from a jam is smaller than the maximal flow.

\begin{figure}[h]
 \centerline{\psfig{figure=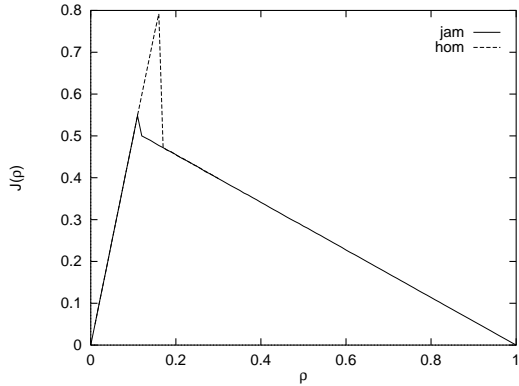,bbllx=50pt,bblly=50pt,bburx=550pt,bbury=400pt,height=5cm}}
\caption{\protect{Fundamental diagram of the BJH model ($v_{max}= 5$,
   $p = 0.01$, $p_s =0.75$) obtained using two different initial conditions,
   namely a completely jammed state (full line) and a homogeneous state
   (broken line).}}
\label{BJH_fund}
\end{figure}

Our simulations show that for $\vm>1$ the overall behaviour of the BJH 
model is very similar to that of the T$^2$ model (for $p_t<1$). 
Therefore we are not going to discuss it here. In Fig.\ \ref{BJH_fund} we
show a typical fundamental diagram.
There is, however, no inflection point and the fundamental diagram
is convex for all $p_s$. This supports the view that the existence
of an inflection point is a 'spatial' effect.

 
\section{NaSch model with velocity-dependent randomization}
\label{sec_nasch}

Here we present a simple generalization of the NaSch model which
incorporates slow-to-start behaviour without introducing memory (like in the
BJH model) or a longer-ranged interaction (like in the T$^2$ model). This
new s2s rule is therefore neither temporal nor spatial.

Instead, a velocity-dependent randomization (VDR) parameter $p=p(v(t))$ is
introduced. This parameter has to be determined {\em before} the
acceleration step R1. For simplicity we here study only the case 
\be
p(v)=\left\lbrace \begin{array}{l@{\quad}l}
p_0 & v=0\\
p   & v>0
\end{array}\right.
\ee
which already contains the most important features of the general case
\cite{barlo}.
Since we are interested in hysteresis phenomena, we restrict ourselves to
the case $p_0\le p$. Note that for $p_0=p$ the NaSch model is recovered.
The cruise-control limit \cite{Pac} corresponds to the choice $p(v_{max})=0$ 
and $p(v)=p$ for $v<\vm$.

In the following we will use a maximum velocity $v_{max}=5$, braking
probability $p=1/64$ of the moving cars and a higher value $p_0 = 0.75$
for the braking probability of stopped cars.  
Simulation runs have been performed for periodic systems with $L=10000$
lattice sites. 

\begin{figure}[h]
  \centerline{\psfig{figure=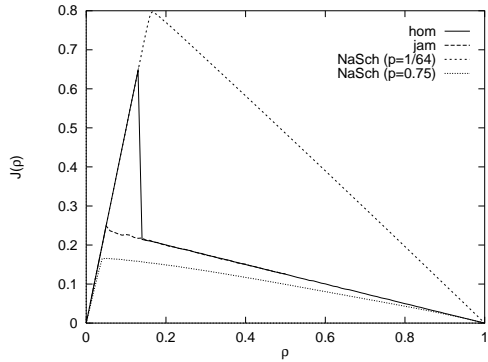,bbllx=40pt,bblly=35pt,bburx=540pt,bbury=410pt,height=5cm}}
\caption{\protect{Fundamental diagram of VDR model ($v_{max}=5$, $p_{0}=0.75$,
$p =1/64$, $L=10000$). For comparison the fundamental diagrams of the
NaSch model with $p=0.75$ and $p=1/64$ are given.}}
\label{fund_nasch}
\end{figure} 

Fig.\ \ref{fund_nasch} shows the fundamental diagram of the modified
model. Obviously the average flow $J(\rho)$ can take two values in the
density interval between $\rho_{1}$ and $\rho_{2}$ depending on the
chosen initialization. The larger values of the average flow can be
obtained using a homogeneous initialization of the system. The lower
branch is obtained starting from a completely jammed state. Moreover,
varying the particle number adiabatically, one can trace a hysteresis
loop. One gets the upper branch by adding cars to the stationary state
with $\rho < \rho_1$ and the lower one by removing cars from the
stationary state with $\rho > \rho_2$.
For a fixed value of $p$, $\Delta J = J(\rho_{2})-J(\rho_{1})$ depends 
linearly on $p_0$ for wide a range of parameters. 

Increasing the system size, we observe a decrease of the density 
$\rho_{2}$ towards the branching density $\rho_1$. 
The jammed states become stable for densities $\rho \geq \rho_1$  
even for global densities very close to $\rho_1$. Again it 
should be noted that the homogeneous states have extremely long life-times
and should therefore be relevant for realistic systems.

It is instructive to compare the fundamental diagram of the VDR model
with those of the corresponding NaSch models. For small densities
$\rho \ll 1$ there are no slow cars in the VDR model since interactions
between cars are extremely rare. Here the flow is given by $J(\rho)
\approx \rho(\vm-p)$, i.e.\ identical to the NaSch model with randomization 
$p$. For large densities $1-\rho \ll 1$, on the other hand, the flow is given
by $J(\rho) \approx (1-p_0)(1-\rho)$ which corresponds to the NaSch model
with randomization $p_0$. For densities close to $\rho=1$, only cars with 
velocities $v_j=0$ or $v_j=1$ exist. The number of moving cars goes to
zero so that asymptotically the flow is completely determined by $p_0$.

The microscopic structure of the jammed states in the VDR model
differs from those found in the NaSch model. While jammed states in
the NaSch model contain clusters with an exponential size-distribution
\cite{duiproc}, one can find phase separation in the VDR model. The reason 
for this behaviour is the reduction of the outflow from a jam. 
If the outflow from a jam is maximal, any small jam in the free flow
regime dissolves immediately since the outflow from such a jam is larger
than the global flow. Therefore phase separation can not occur in that
case. However, if the outflow from a jam is reduced,
the density in the free flow regime is smaller than the density of
maximum flow and cars can propagate freely in the low density part of
the lattice. Due to the reduction of the density in the free flow
regime, no spontaneous formation of jams is observable in the stationary
state, if fluctuations in the free flow regime are rare. 

This picture is supported by a simple phenomenological approach. Obviously 
the flow in the homogeneous branch is given by $J_{hom}=\rho(v_{max}-p)=
\rho v_{f}$, because every car can move with the free-flow velocity $v_{f}$. 
Assuming that the high density states
are phase separated, we can obtain the second branch of the fundamental
diagram. The phase separated states consist of a large jam and a free
flow regime, where each car moves with velocity $v_f$. The density in
the free flow regime $\rho_f$ is determined by the average waiting
time $T_w = \frac{1}{1-p_0}$  of the first car in the jam  and $v_f$, because
neglecting interactions between cars, the average distance of two
consecutive cars is given by $\Delta x = T_w v_{f} + 1 = \rho_f^{-1}$. Using
the normalisation $L = N_J + N_F \Delta x$ ($N_{F(J)}$ is the number of cars
in the free flow regime (jam)) we find that the flow is given by 
\begin{equation}
  \label{jjam}
  J_{sep}(\rho) = (1-p_0)(1-\rho).
\end{equation}
Obviously $\rho_f$ is precisely the lower branching density $\rho_1$, because
for densities below $\rho_f$ the jam-length is zero.
It should be noted that this approach is only valid for $p_0 \gg p$ and
$v_{max} >1$. The condition $p\ll 1$ guarantees that interactions of cars
due to velocity fluctuations are rare. As a consequence, the jam is
compact in that limit. For increasing $p$, the jam becomes less dense.
In the case $\vm=1$, cars can stop spontaneously, even in the free-flow regime.
If $p_0$ is large enough, these cars might initiate a jam. This is the
basic reason why hysteresis is usually not observed for $\vm=1$.

\begin{figure}[h]
 \centerline{\psfig{figure=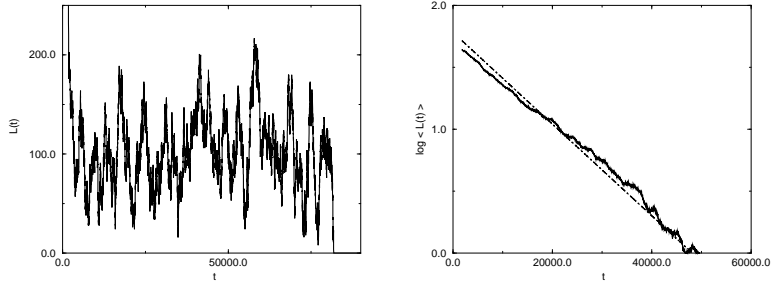,bbllx=0pt,bblly=0pt,bburx=1170pt,bbury=530pt,height=5cm}}
 \caption{\protect{Time-dependence of the length $L_{Jam}(t)$ of the jam 
     for $\rho=0.095$. The left part of the figure shows the time evolution of
     the length $L_{Jam}(t)$ of one sample. The average $\langle L_{Jam}(t)
     \rangle$ over 10000 samples (right part of the figure)
     shows an exponential decay.}}
\label{jam_length}
\end{figure}

Measurements of the average flow show that the lower branch of the
fundamental diagram is not stable near the density $\rho_1$, if small
system sizes are considered. Therefore we performed a more 
detailed stability analysis of the homogeneous and
the jammed state near $\rho_1$ and  $\rho_2$.  Close to $\rho_1$, the
large jam present in the initial configuration dissolves and the average length
$\langle L_{Jam}(t)\rangle$ decays exponentially in time 
(Fig.~\ref{jam_length}). 
It should be noted that this behaviour is not the consequence of a 
continuous "melting" of the large jam. In contrast, the jam-length 
$L_{Jam}(t)$ is strongly fluctuating without any obvious systematic 
time-dependence (Fig.~\ref{jam_length}).  
Once a homogeneous state without a jammed car is reached, no new jams are
formed. Therefore the homogeneous state is stable near $\rho_1$.
For large system sizes the jammed states
are stable for $\rho \geq \rho_1$, also for densities only slightly
above $\rho_1$, because the average length of the jam is
proportional to the system size, while the fluctuations grow
sub-extensive. 
\begin{figure}[h]
  \centerline{\psfig{figure=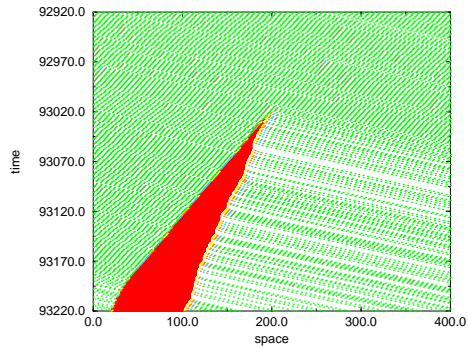,bbllx=0pt,bblly=0pt,bburx=570pt,bbury=530pt,height=6cm}}
  \caption{\protect{Space-time diagram of the VDR model for
    $\rho=0.15, L=400, p =0.01 $ and $p_0=0.5$. The homogeneous initial
      state is not destroyed immediately, but after approximately 93000 
      lattice updates. In the outflow regime of the jam the density is 
      reduced compared to the average density.}}
\label{st_diag}
\end{figure}

Analogous to the metastable jammed states near $\rho_1$, homogeneous
initializations for densities slightly above $\rho_2$  lead to
metastable homogeneous states with short lifetimes. Fig.~\ref{st_diag}
shows the spontaneous formation of jams due to velocity fluctuations. The
finite lifetimes of the homogeneous states are the qualitative
difference between this model and the cruise-control limit \cite{Pac} of
the NaSch model, where the time evolution of homogeneous states at low 
densities is completely deterministic.

For large system sizes the density difference $\Delta \rho = \rho_2 -\rho_1$ 
decreases. This can be explained by looking at the
mechanism of an emerging jam. Jams emerge due to velocity fluctuations
in dense regions of the homogeneous states, where the distance $d_j+1$ 
between consecutive cars is less than $v_{max}$. 
In these regions all following cars have to slow down if a car in front
breaks in the randomization step. Over-reactions of following cars
finally can cause jams. The probability to find clusters of an
appropriate length for a given density is proportional to the system
size. Therefore we expect that all homogeneous states are unstable for
$\rho > \rho_1$ in the thermodynamic limit.

\begin{figure}[h]
  \centerline{\psfig{figure=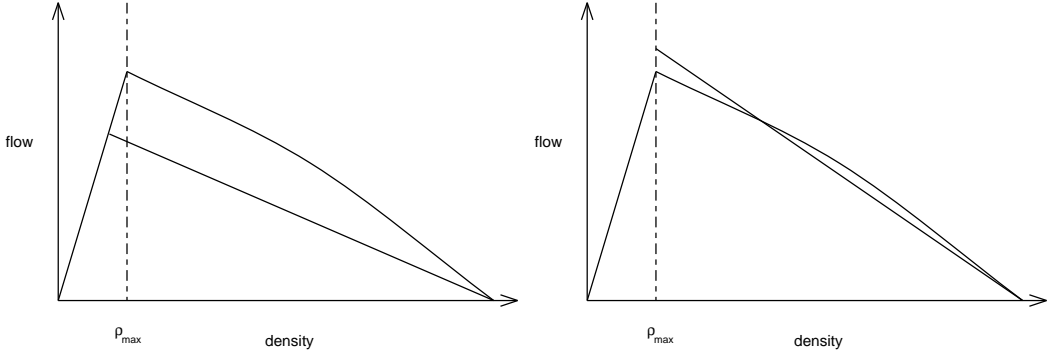,bbllx=15pt,bblly=560pt,bburx=585pt,bbury=760pt,height=5cm}}
  \caption{\protect{Illustration of the flow branches a) in the
presence of phase separation and b) without phase separation.}}
\label{argument}
\end{figure}

\begin{figure}[h]
  \centerline{\psfig{figure=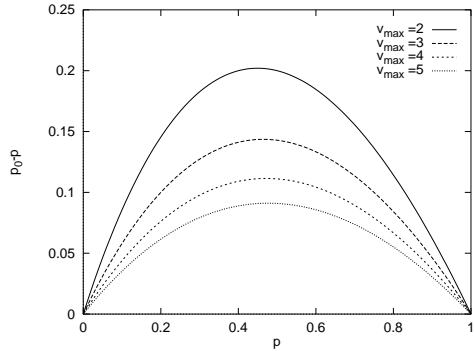,bbllx=40pt,bblly=30pt,bburx=550pt,bbury=415pt,height=5cm}}
  \caption{\protect{Estimates for the minimum difference $p_0 -p$ necessary
to observe phase separation.}}
\label{fig_comp}
\end{figure}

For higher values of $p$ the lifetime of homogeneous states at densities
$\rho > \rho_1$ is very small. Nevertheless one can observe the same
microscopic structure of the high density states as long as the outflow
of a jam is sufficiently reduced. A {\em rough} estimate for the
minimal difference between $p_0$ and $p$ necessary to observe phase separation
can be obtained from the following arguments based on the features 
of a typical fundamental diagram of the VDR model at small $p$ and
large $\vm$ (see Fig.\ \ref{argument}).
In the following we assume that the fundamental diagram is perfectly linear
up to $\rho_{2}$, i.e.\ $J_{hom}(\rho) = \rho(\vm-p)$. As mentioned before,
the phase-separated branch is described by $J_{sep}(\rho) = (1-p_0)(1-\rho)$ 
(see Eq.~(\ref{jjam})). In order to find phase separation, the flow in the 
phase-separated branch at density $\rho_{2}$ must be lower than the flow in 
the homogeneous branch, i.e.\ $J_{hom}(\rho_{2}) > J_{sep}(\rho_{2})$
(left part of Fig.\ \ref{argument}). For a situation as depicted in the
right part of Fig.\ \ref{argument} (i.e.\ for  $J_{hom}(\rho_{2}) <
J_{sep}(\rho_{2})$), no phase-separation would be found.
In the following we approximate $\rho_2$ by $\rho_{max}$, the maximum of
the corresponding NaSch model with randomization $p$. Using the estimate
$\rho_{max}=(1-p)/(v_{max}+1)$ \cite{eisi}, the inequality for the current
at $\rho_2\approx \rho_{max}$ can be used to obtain an estimate for the 
minimal difference $p_0-p$ necessary to observe phase separation (see 
Fig.~\ref{fig_comp}).
Near the deterministic limits ($p=0$ and $p=1$) and for large values of 
$v_{max}$, already a small difference $p_0-p$ suffices to generate
phase separation.
For all nondeterministic cases the states of maximum flow are not completely
homogeneous and the flow in the high density branch is somewhat lower
than $J_{sep}(\rho) = (1-p_0)(1-\rho)$ for larger values of
$p$. In addition, the large jam is not compact anymore.
 
In order to substantiate the picture developed in this section we also
investigated correlation functions and the effect of perturbations 
(i.e.\ stopping a car temporarily to induce a jam) \cite{RB_dipl}. 
The results are in full agreement with our interpretation and will be 
presented elsewhere \cite{BSSS}.

Another interesting result concerns the outflow from a jam.
Our simulations show \cite{BSSS} that the outflow is (almost)
independent of the density of cars and depends only on $p$. 
This in agreement with measurements on real traffic 
\cite{KernerReh2} where it was found that the outflow only depends on 
road and weather conditions and typical characteristics of the cars.
Furthermore we have found that a structure consisting of two
separated jams is rather stable and can exist for long times. This
is also in agreement with experimental observations \cite{KernerReh2}.

For $\vm=1$ we used the so-called car-oriented mean-field theory \cite{comf}
in order to obtain an analytical description of the fundamental diagram.
The results are in good agreement with simulations \cite{RB_dipl}. As
expected, one finds no phase separation at any $p_0$ for $p>0$. In the
deterministic limit, however, a hysteresis curve can be observed for 
$p_0>0$. The fundamental diagram constists of two linear branches and is 
in perfect agreement with simulations \cite{BSSS}.


\section{Summary and Discussion}
\label{sec_disc}

The ability to describe the existence of hysteresis is a serious test
for any traffic flow model. In this paper we have investigated
several variants of the NaSch model and showed numerically the
existence of metastable states in a density region close to the
maximum flow.

The hysteresis in the fundamental diagram is related
to the existence of metastable states. The latter are a
consequence of phase separation in a certain density regime.
The reason for this phase separation, on the other hand, is
a reduced outflow from jams which destabilizes clusters forming
in the outflow region. Such a reduction can be
incorporated most easily by introducing slow-to-start rules
which try to model the restart behaviour of standing vehicles
more realistically. The three models presented here used
different s2s rules which can be classified as spatial, temporal
and velocity-dependent, respectively. For maximum velocity $\vm>1$
the models exhibit a similar behaviour (see below). 
For $\vm=1$ the spatial s2s rule of the T$2$ model
is exceptional since it leads to a existence of a phase transition
to a completely jammed state in the limit $p_t\to 1$ where $p_t$
is the s2s parameter.

The NaSch model with velocity-dependent randomization shows the
coexistence of phase separated and homogeneous states in a density
interval near the density $\rho_2$ of maximum flow. Near $\rho_2$
interactions between cars become important and one can find
spontaneous formation of jams. Therefore the reduction of the density
in the outflow regime of a jam leads to stable phase separated states.
The reduction of interactions between cars in the free flow regime can
be confirmed by a phenomenological approach, which gives very accurate
results. In contrast to the cruise-control limit of the NaSch model
\cite{Pac}, where also metastable states can be found, fluctuations
are present in both coexisting states. 

The results presented here have interesting applications. In real traffic
one is usually interested in stabilizing the homogeneous branch of
the fundamental diagram in order to maximize the throughput. This can
be done using signals to control the inflow, as in the case of the
Lincoln tunnel in New York \cite{lincoln}. Results of simulations
corresponding to such a situation will be presented in a future
publication \cite{BSSS}.

In \cite{Krauss} a family of space-continuous models has been
investigated. Depending on the values of the maximum acceleration and
deceleration three classes can be distinguished. Class III shows no
realistic behaviour since no spontaneous jams are formed. A continuous
analogue of the NaSch model belongs to class II (high decelaration limit).
Here one finds spontaneous formation of jams, but no metastable states.
Finally, in class I spontaneous jamming as well as metastable states
are found. On a macroscopic level, models in classes I and II can be
distinguished by the ordering of the densities $\rho_f$, the density
of the outflow from a jam, and $\rho_c$, where the density of a
homogeneous flow becomes unstable\footnote{Often it is not easy to
define these quantities properly, especially in the  presence of noise.}. 
The discrete models discussed here have very simple braking rules, but
nevertheless exhibit the behaviour of class III.
The existence of metastable states is therefore not related to the use 
of continuous space coordinates or ``realistic'' braking rules 
\cite{zprmod1,zprmod2} (e.g.\ anticipation of the behaviour of the 
driver ahead).  

Already the NaSch model has a tendency towards the formation
of metastable states. The outflow from a megajam is maximal only in the
deterministic case $p=0$ and in the so-called cruise-control limit
\cite{Pac}. In the generic case $p>0$ the outflow is not maximal 
\cite{eisidipl}, but organizes itself towards the `critical' current 
$J_c$ where the correlation length becomes maximal \cite{eisi}.
Since the difference between $J_{max}$ and $J_c$ is rather small,
it is very difficult to observe metastable states in the
NaSch model.

The considerations in this paper show the flexibility of the CA
approach to traffic flow problems. A rather simple and natural
extension of the rules of the NaSch model allows us to describe
the formation of metastable states in the fundamental diagram.
The introduction of a velocity-dependent randomization $p(v)$ makes
it possible to control the properties of the free flow and congested
flow independently. Experimentally it has been found \cite{KernerReh2}
that for real traffic the reduction of the outflow $J_{out}$ compared
to the maximum flow $J_{max}$ is approximately $J_{max}/J_{out}\approx 1.5$.
This value can be used to determine a realistic value of $p_0-p$.

If one is willing to give up an overly realistic description of the
interactions between the vehicles one can obtain rather simple CA
models capable of describing even the fine-structure of traffic
flow in a satisfactory way.

%
%


\begin{thebibliography}{55}
%
\bibitem{NaSch} 
  K.\ Nagel, M.\ Schreckenberg:
  J. Physique {\bf I2}, 2221 (1992)

\bibitem{juelich} D.E.\ Wolf, M.\ Schreckenberg, A.\ Bachem (Eds.):
{\em Traffic and Granular Flow}, World Scientific (1996)

\bibitem{duis} M.\ Schreckenberg, D.E.\ Wolf (Eds.):
{\em Traffic and Granular Flow '97}, Springer (1998)

\bibitem{helbing} D.\ Helbing: {\em Verkehrsdynamik}, Springer (1997)

\bibitem{may} A.D\ May: {\em Traffic Flow Fundamentals}, Prentice Hall (1990)

\bibitem{dallas} M.\ Rickert, K.\ Nagel: Int.\ J.\ Mod.\ Phys. {\bf C8}, 
483 (1997)\\
K.\ Nagel, C.L.\ Barrett: Int.\ J.\ Mod.\ Phys. {\bf C8}, 505 (1997)

\bibitem{duisburg} J.\ Esser, M.\ Schreckenberg: Int.\ J.\ Mod.\ Phys.\
{\bf C8}, 1025 (1997)

\bibitem{zprmod1}  S.\ Krau{\ss}, P.\ Wagner, C.\ Gawron:
  Phys. Rev. {\bf E54}, 3707 (1996) 

\bibitem{zprmod2}
  S.\ Krau{\ss}, P.\ Wagner, C.\ Gawron:
  Phys. Rev. {\bf E55}, 5597 (1997) 

\bibitem{KK} B.\ Kerner, S.L.\ Klenov, P.\ Konh\"auser: Phys.\ Rev.\ {\bf 56},
4200 (1997)

\bibitem{bando} M.~Bando, K.~Hasebe, A.~Nakayama, A.~Shibata, Y.~Sugiyama:
Phys.\ Rev.\ {\bf E51}, 1035 (1995)

\bibitem{kikuchi} S.\ Yukawa, M.\ Kikuchi: J.\ Phys.\ Soc.\ Jpn.\ {\bf 64},
35 (1995)

\bibitem{helbinggas} D.\ Helbing: Phys.\ Rev.\ {\bf 53}, 2366 (1996)

\bibitem{Treit} J.\ Treiterer: Ohio State Technical Report No.\ PB 246 094
(1975)

\bibitem{Kerner2}  B.S.\ Kerner, H.\ Rehborn: Phys.~Rev.~Lett.~{\bf 79}, 4030
(1997)

\bibitem{helbingexp} D.\ Helbing: Phys.\ Rev.\ {\bf E55}, R25 (1997)

\bibitem{KernerReh2}  B.S.\ Kerner, H.\ Rehborn:
  Phys.\ Rev.\ {\bf E53}, R1297 (1996)

\bibitem{slow_comf} 
  A.\ Schadschneider, M.\ Schreckenberg:
  Ann.\ Physik {\bf 6}, 541 (1997) 

\bibitem{T2mod}   M.\ Takayasu, H.\ Takayasu:
  Fractals {\bf 1}, 860 (1993)

\bibitem{fukui} M.\ Fukui, Y.\ Ishibashi: J.\ Phys.\ Soc.\ Jpn.\ {\bf 66},
385 (1997) 

\bibitem{BJH} S.C.\ Benjamin, N.F.\ Johnson, P.M.\ Hui: J.\ Phys.\ 
{\bf A29}, 3119 (1996)

\bibitem{Krauss}  S.\ Krau{\ss}: in \cite{duis}

\bibitem{duiproc}  A.\ Schadschneider: in \cite{duis}

\bibitem{Pac} K.\ Nagel, M.\ Paczuski:
  Phys. Rev. {\bf E51}, 2909  (1995)


\bibitem{barlo} R.\ Barlovic, L.\ Santen, A.\ Schadschneider, 
M.\ Schreckenberg: in \cite{duis}

\bibitem{eisi} B.\ Eisenbl\"atter, L.\ Santen, A.\ Schadschneider, 
   M.\ Schreckenberg: 
   Phys.\ Rev.\ {\bf E57}, 1309 (1998)

\bibitem{RB_dipl} R.\ Barlovic: Diploma Thesis, Universit\"at Duisburg (1998)

\bibitem{BSSS} R.\ Barlovic, L.\ Santen, A.\ Schadschneider, 
M.\ Schreckenberg: in preparation


\bibitem{comf} A.\ Schadschneider, M.\ Schreckenberg: J.~Phys.\ {\bf A30}, 
L69 (1997)

\bibitem{lincoln} H.\ Greenberg, A.\ Daou: Operations Res.~{\bf 8}, 524 (1960)

\bibitem{eisidipl} B.\ Eisenbl\"atter: Diploma Thesis, Universit\"at
Duisburg (1996)

\end{thebibliography}
\end{document}